\documentclass[aps,prl,showpacs,twocolumn]{revtex4}
\usepackage{epsfig}
\usepackage{amsfonts}
\begin{document}

%%%%%%%%%%%%%%%%%%%%%%%%%%%%%%%%%%%%%%%%%%%%%%%%%%%%%%%%%%%%%%%%%%%%%
%%%%%%%%%%%%%%%%%%%%%         Title       %%%%%%%%%%%%%%%%%%%%%%%%%%%
%%%%%%%%%%%%%%%%%%%%%%%%%%%%%%%%%%%%%%%%%%%%%%%%%%%%%%%%%%%%%%%%%%%%%

\title{Overspinning a nearly extreme charged black hole via a quantum tunneling process}

%%%%%%%%%%%%%%%%%%%%%%%%%%%%%%%%%%%%%%%%%%%%%%%%%%%%%%%%%%%%%%%%%%%%%
%%%%%%%%%%%%%%%%%%%%     Authors & Addresses  %%%%%%%%%%%%%%%%%%%%%%%
%%%%%%%%%%%%%%%%%%%%%%%%%%%%%%%%%%%%%%%%%%%%%%%%%%%%%%%%%%%%%%%%%%%%%

\author{George E. A. Matsas and Andr\'e R. R. da Silva}
\address{Instituto de F\'\i sica Te\'orica, Universidade Estadual Paulista,
         Rua Pamplona 145, 01405-900, S\~ao Paulo, S\~ao Paulo,
         Brazil}

\date{\today}

%%%%%%%%%%%%%%%%%%%%%%%%%%%%%%%%%%%%%%%%%%%%%%%%%%%%%%%%%%%%%%%%%%%%%%
%%%%%%%%%%%%%%%%%%%           Abstract            %%%%%%%%%%%%%%%%%%%%
%%%%%%%%%%%%%%%%%%%%%%%%%%%%%%%%%%%%%%%%%%%%%%%%%%%%%%%%%%%%%%%%%%%%%%

\begin{abstract}
We examine a nearly extreme macroscopic Reissner-Nordstrom black
hole in the context of semiclassical gravity. The absorption rate
associated with the quantum tunneling process of scalar particles
whereby this black hole can acquire enough angular momentum to
violate the weak cosmic censorship conjecture is shown to be
nonzero.

\end{abstract}


\maketitle

%%%%%%%%%%%%%%%%%%%%%%%%%%%%%%%%%%%%%%%%%%%%%%%%%%%%%%%%%%%%%%%%%%%%%%%
%%%%%%%%%%%%%%%%%%%%%         Text Body          %%%%%%%%%%%%%%%%%%%%%%
%%%%%%%%%%%%%%%%%%%%%%%%%%%%%%%%%%%%%%%%%%%%%%%%%%%%%%%%%%%%%%%%%%%%%%%

The {\em weak cosmic censorship conjecture} (WCCC) plays a
fundamental role in black hole physics~\cite{grw,grqc,clarke}. The
validity of the WCCC is a necessary condition to ensure the
predictability of the laws of physics~\cite{hawk}. This conjecture
asserts that spacetime singularities coming from the complete
gravitational collapse of a body must be encompassed by the event
horizon of a black hole. We show that quantum effects may challenge
the WCCC.

General Relativity is the most successful spacetime theory currently
available~\cite{grw,lsss}. It has provided us with much insight on
the WCCC. According to the uniqueness theorems~\cite{nohair}, all
stationary black hole solutions of Einstein-Maxwell equations are
uniquely determined by three conserved parameters: the gravitational
mass $M$, the electric charge $Q$, and the angular momentum $J$,
which satisfy
\begin{equation}
M^2 \geq Q^2+(J/M)^2. \label{ccc}
\end{equation}
In particular, $M^2 = Q^2+(J/M)^2$ characterizes extreme black
holes. (See Ref.~\cite{farrugia} for a possible way to generate
them.) We note that $M^2 < Q^2+(J/M)^2$ are associated with naked
singularities rather than black holes. (We assume natural units,
$c=G=\hbar=1$, unless stated otherwise.) A number of
{\it{classical}} results have given full support to the WCCC.
Stationary black holes are stable under linear
perturbations~\cite{vish,price,kay,whit} what yields a good testing
ground for this conjecture. By analyzing the time development of a
thin charged shell with positive proper mass, Boulware concluded
that its total collapse does not lead to a naked
singularity~\cite{shell}. Next, Wald considered an extreme black
hole and wondered whether or not it would be possible to slightly
overcharge and/or overspin it in order to violate the
WCCC~\cite{gedan}. Wald shows that when the hole swallows a
{\it{classical}} rotating and/or charged test particle, its mass $M$
increases enough to compensate any extra gain of angular momentum
and/or charge during the process such that the WCCC is upheld. (See
also Ref.~\cite{tristan}.) Even though no one has demonstrated the
validity of the WCCC in the context of General Relativity yet, it is
quite possible that this is achieved sometime in the
future~\cite{grqc}.

On the other hand, it is well known that when one takes into account
quantum effects some classical results associated with black holes
can be deeply modified. For instance, the classical theory says that
the area of a black hole does never decrease under any processes,
whereas quantum mechanics shows that it will be reduced as a
consequence of the black hole radiance~\cite{hawk_rad}. This
celebrated phenomenon which dwells on the theory of quantum fields
in curved spacetimes~\cite{birrel,fulling} triggered a broad
interest on the issue of black hole thermodynamics (see
Refs.~\cite{bht,lrr} and references therein). It seems natural to
inquire then as to whether quantum mechanics can affect WCCC as
well. An investigation in these lines was recently performed by Ford
and Roman~\cite{ford1,ford2,ford3} who analyzed the possibility of
violating the WCCC by injecting negative energy into an extreme
charged black hole, i.e. $Q=M$ and $J=0$. They concluded that the
positive energy flux which always follows the negative one would
preserve the WCCC.

This {\it{Letter}} revisits the question of whether or not the WCCC
remains valid when one includes quantum effects from a different
perspective. We consider a nearly extreme macroscopic
Reissner-Nordstrom black hole, $Q/M \lesssim 1$, and analyze a
quantum tunneling process whereby the black hole can acquire enough
angular momentum to overspin, $M^2 < Q^2+(J/M)^2$, and therefore
violate the WCCC. This is treated in the context of a well
established semiclassical approach, where the low-energy sector of a
free massless scalar field is canonically quantized outside the
black hole~\cite{rapid,cm} (see also Ref.~\cite{chm}). We stress 
that we only deal with low-frequency modes in order to minimize 
backreaction effects. The absorption rate of scalar particles is 
calculated and used to discuss in what regime the WCCC could be 
violated by this quantum tunneling process. We eventually comment 
on the possible limitations of the semiclassical approach and 
delineate some aspects that should be present after a full quantum 
gravity analysis is performed.

The line element of a Reissner-Nordstrom black hole can
be written in the form~\cite{grw}
\begin{equation}
ds^2=f(r)dt^2-f^{-1}(r)dr^2-r^2(d\theta^2+\sin^2\theta \;d\phi^2),
\label{rn}
\end{equation}
where $f(r)=(1-r_+/r)(1-r_-/r)$ and $r_\pm = M \pm \sqrt{M^2-Q^2}$.
The outer event horizon is located at $r = r_+$. This black hole is
assumed to be macroscopic (i.e. $M$ must be much larger than the
Planck mass $M_{\rm{P}}$) in order to guarantee the applicability of
the Semiclassical Gravity Theory.

We analyze now the process in which massless scalar particles are
beamed from far away towards the black hole. Because of the
existence of an effective scattering potential, low-energy
particles, $\omega \approx 0$, are mostly reflected back to
infinity. As a matter of fact, the {\it{few}} particles which enter
into the hole must quantum-mechanically tunnel into it. We will only
focus on soft particles because in this case, backreaction effects
which are in general present~\cite{hiscock}, should be significantly
dumped by assuming large enough black holes ($M \gg \hbar \omega /
c^2$). In order to maximize the tunneling probability it is
convenient to consider nearly extreme black holes which may overspin
by the absorption of a single particle. We also note that in order
to disregard any discharge of the hole by electron-positron pair
production at the horizon vicinity~\cite{gibbons,ford1} one must
assume black holes with mass $M > 10^{5} M_{\odot}$, where
$M_{\odot}$ is the solar mass. As a byproduct of these requirements,
the Hawking temperature associated with these holes, which goes like
$\sqrt{M^2-Q^2}/r^{2}_{+}$, is negligible.

\begin{figure}[t]
\epsfig{file=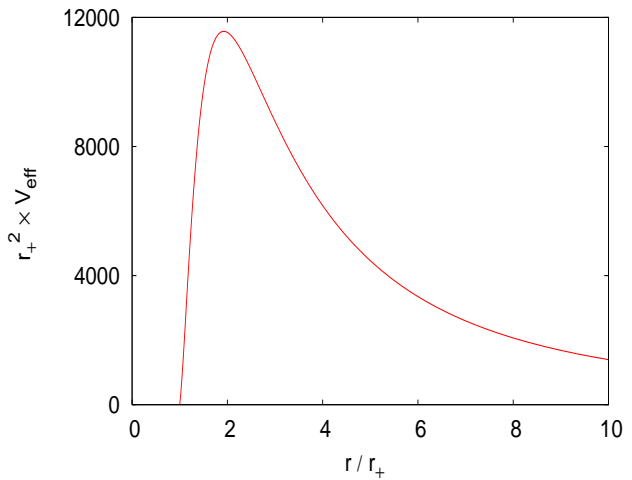,angle=0, width=8.5cm, height=6.5cm}
\caption{The effective scattering potential $V_{\rm{eff}}$ is plotted for $M=10^2 M_{\rm{P}}$,
$l=413$ and $Q=M-e$, where $e$ is the electron charge. 
(The numerical values were taken from Fig.~\ref{f2}.)
\label{f3}}
\end{figure}

The dynamics of a free massless real scalar field $u_{\lambda \omega
l m}$ is described by the usual Klein-Gordon equation
$\nabla^{\mu}\nabla_{\mu}u_{\lambda \omega l m}=0$. By using the
ansatz
\begin{equation}
u_{\lambda \omega l m}(t,r,\theta,\phi)=
\sqrt{\frac{\omega}{\pi}}\frac{\psi_{\lambda \omega l}(r)}{r} Y_{l
m}(\theta,\phi)e^{-i\omega t} \label{ansatz}
\end{equation}
with $\omega \geq 0$, $l \geq 0$ ($l \in \mathbb{N}$) and $m \in
[-l,l]$, we reduce the Klein-Gordon equation to an one-dimensional
differential equation for $\psi_{\lambda \omega l}$:
\begin{equation}
\left[-f(r)\frac{d}{dr}\left(f(r)\frac{d}{dr}\right)+ V_{\rm
eff}(r)\right]\psi_{\lambda \omega l}(r)= \omega^2\psi_{\lambda
\omega l}(r), \label{dre1}
\end{equation}
where
\begin{equation}
V_{\rm eff}(r) =
f(r)\left[{l(l+1)}/{r^2} + {2M}/{r^3} - {2Q^2}/{r^4}\right]
\label{esp}
\end{equation}
is the effective scattering potential (see Fig.~\ref{f3}).
Eq.~(\ref{dre1}) possesses two sets of independent solutions
associated with purely incoming modes (i) from the white-hole
horizon ${\cal{H}}^-$, $\lambda = \rightarrow$, and (ii) from the
past null infinity ${\cal{I}}^-$, $\lambda = \leftarrow$.

Thus, the corresponding scalar field operator $\Phi(x)$ can be
expanded using the complete set of normal modes as
\begin{equation}
\Phi (x)=\sum^{\leftarrow}_{\lambda=\rightarrow}\sum^{\infty}_{l=0}\sum^{l}_{m=-l}
\int^{\infty}_{0}d\omega\left[ a_{\lambda \omega l m}u_{\lambda
\omega l m}(x)+ {\rm H.c.} \right], \label{srf}
\end{equation}
where $u_{\lambda \omega l m}(x)$ are orthonormalized according to
the Klein-Gordon inner product~\cite{birrel}. With this
normalization, the operators $a_{\lambda \omega l m}$ and
$a^{\dagger}_{\lambda \omega l m}$ satisfy the usual commutation
relations
$$
[a_{\lambda \omega l m},
a^{\dagger}_{\lambda^{\prime}\omega^{\prime} l^{\prime}m^{\prime}}]=
\delta_{\lambda \lambda^{\prime}} \delta_{l l^{\prime}} \delta_{m
m^{\prime}} \delta (\omega-\omega^{\prime}).
$$
The vacuum state $|0\rangle$ is defined as $ a_{\lambda \omega l
m}|0\rangle=0|0\rangle$ for every $\lambda$, $\omega$, $l$ and
$m$~\cite{boulware}.

It is well known that the solutions of Eq.~(\ref{dre1}) cannot be
trivially written in terms of the usual special functions. (For more
details, see Ref.~\cite{candelas}.) Nevertheless, this is not so in
the low-energy regime. In fact, by making $\omega = 0$ we can recast
Eq.~(\ref{dre1}) as a Legendre's equation~\cite{arfken}:
\begin{equation}
\frac{d}{dv}\left[(1-v^2)\frac{d}{dv}\left(\frac{\psi_{\lambda \omega
l}(\bar{r}[v])}{\bar{r}[v]}\right)\right]+
l(l+1)\frac{\psi_{\lambda \omega l}(\bar{r}[v])}{\bar{r}[v]}=0 ,
\label{der2}
\end{equation}
where $v \equiv (2\bar{r}-1)/(\bar{r}_+ -\bar{r}_-)$ is a
dimensionless variable with $\bar{r}=r/2M$ and
$\bar{r}_{\pm}=r_{\pm}/2M$. Since we are interested here in
particles coming from infinity, we will only write the radial part
corresponding to the normal mode $\lambda = \leftarrow$:
\begin{equation}
\psi_{\leftarrow \omega l}(\bar{r})= C_{\omega l} \bar{r}
P_l[v(\bar{r})].\label{lp}
\end{equation}
It is worthwhile to note that $P_l[v(\bar{r})]$ behaves as $P_{l}(v)
\approx [(2l)! / 2^l (l!)^2] v^{l}$ for $v \gg 1$ (i.e. $r \gg
r_{+}$). Here $C_{\omega l}$ is a normalization constant to be
determined. For this purpose,  the behavior of the radial mode 
functions near the horizon and at large distances for any value 
of $\omega$ is investigated~\cite{cm}. In lieu of dealing with 
Eq.~(\ref{dre1}) we introduce the Regge-Wheeler dimensionless coordinate
$$
r^{\ast} \equiv
\bar{r}+\frac{\bar{r}^{\;2}_{+}\ln|\bar{r}-\bar{r}_{+}|-
\bar{r}^{\;2}_{-}\ln|\bar{r}-\bar{r}_{-}|}{\bar{r}_{+}-\bar{r}_{-}}
$$
to transform it in a Schr\"odinger-like equation:
\begin{equation}
\left[ -\frac{d^2}{d r^{\ast \; 2}} + 4M^2 V_{\rm eff}
[r(r^{\ast})]\right]\psi_{\leftarrow \omega l}(r^{\ast}) =
4M^2\omega^2\psi_{\leftarrow \omega l}(r^{\ast}). \label{der3}
\end{equation}
At large distances from the event horizon, $r^{\ast} \gg 1$, the
effective scattering potential falls off approximately as
$l(l+1)/r^2$. This implies that the purely incoming modes from the
past null infinity can be written as
\begin{eqnarray}
&& \psi_{\leftarrow \omega l}(r^{\ast}) \approx 
(-i)^{l+1} M r^{\ast} h_l^{(2)}(2M\omega r^{\ast}) 
\nonumber \\
&& + {i^{l+1}}{\cal R}_{
\omega l} M r^{\ast} h_l^{(1)} ( 2M\omega r^{\ast}) \;; 
r^{\ast} \gg 1,
\label{tail2}
\end{eqnarray}
where $h_l^{(j)}(2M\omega r^{\ast})$, with $j=1,2$, are the
spherical Hankel functions. Next, we recall that near the event
horizon, $r^{\ast} <0$ and $|r^{\ast}|\gg 1$, the potential barrier
is very small: $V_{\rm eff} \approx 0$. Thus, the purely incoming
modes take the form
\begin{equation}
\psi_{\leftarrow \omega l}(r^{\ast}) \approx (2 \omega)^{-1}{\cal
T}_{ \omega l}\; e^{-2i M\omega r^{\ast}} \;; r^{\ast} <0,
|r^{\ast}|\gg 1. \label{tail1}
\end{equation}
Here $|{\cal R}_{ \omega l}|^2$ and $|{\cal T}_{ \omega l}|^2$ stand
for the reflection and transmission coefficients, respectively, and
satisfy the usual probability conservation equation $|{\cal R}_{
\omega l}|^2 + |{\cal T}_{ \omega l}|^2=1$. The normalization of
Eqs.~(\ref{tail2}) and (\ref{tail1}) were accomplished by means of
the Klein-Gordon inner product in agreement with the convention
fixed below Eq.~(\ref{srf}).

In order to determine $C_{\omega l}$ and ${\cal{T}}_{\omega l}$ we
need to match the radial functions $\psi_{\leftarrow \omega
l}(\bar{r})$ and $\psi_{\leftarrow \omega l}(r^{\ast})$ given in
Eqs.~(\ref{lp}) and (\ref{tail2})-(\ref{tail1}), respectively.
Firstly, let us write Eq.~(\ref{tail2}) in the low-energy regime,
i.e. $2M\omega r^{\ast} \ll 1$:
\begin{equation}
\psi_{\leftarrow \omega l}(r^{\ast}) \approx
\frac{(-i)^{l+1}2^{2l+1}(l!)\omega^l(Mr^{\ast})^{l+1}}{(2l+1)!}.\label{edrap}
\end{equation}
We note here that ${\cal R}_{ \omega l} \approx (-1)^{l+1} +
{\cal{O}}(\omega)$. Physically, this means that most incoming
particles from the past null infinity will be reflected back by the
potential barrier. Now, the asymptotic expansion of Eq.~(\ref{lp})
leads to
\begin{equation}
\psi_{\leftarrow \omega l}(\bar{r}) \approx C_{\omega l}
\frac{(2l)!\;\bar{r}^{\;l+1}}{(l!)^2(\bar{r}_+ -\bar{r}_-)^l}.
\label{lpap}
\end{equation}
Then, by comparing Eqs.~(\ref{edrap}) and (\ref{lpap}), one obtains
\begin{equation}
C_{\omega l}=
(-i)^{l+1}\frac{2^{2l+1}(l!)^{3}(\bar{r}_+-\bar{r}_-)^{l}M^{l+1}\omega^l}{(2l)!(2l+1)!}.\label{cte}
\end{equation}
In what follows, $|{\cal T}_{ \omega l}|^2$  can be determined by
fitting Eqs.~(\ref{tail1}) and (\ref{lpap}):
\begin{equation}
|{\cal T}_{ \omega l}|^2 = \frac{2^{4l+4}(\bar{r}_+)^2(\bar{r}_+ -
\bar{r}_-)^{2l}(l!)^6 M^{2l+2}\omega^{2l+2}} {[(2l+1)! (2l)!]^2},
\label{T}
\end{equation}
where we recall that $\bar{r}_\pm = r_\pm / 2M$ and $r_\pm = r_\pm (M,Q)$ 
is defined below Eq.~(\ref{rn}).
{\it{This is the absorption rate associated with the scalar
particles which tunnel the potential barrier towards the hole}}.

\begin{figure}[t]
\epsfig{file=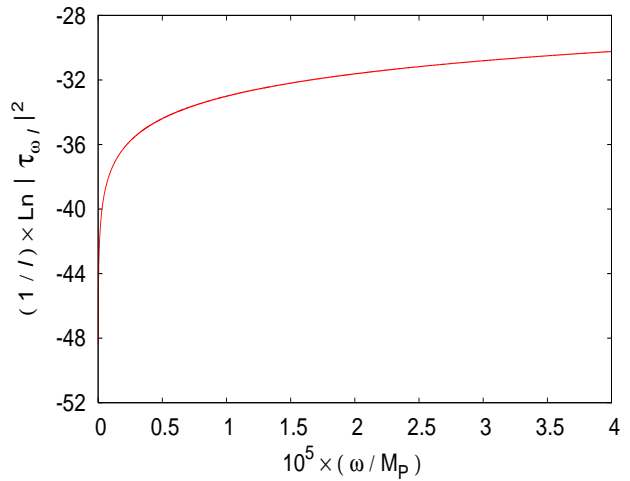,angle=0, width=8.5cm, height=6.5cm}
\caption{The absorption probability  $|{\cal{T}}_{\omega l}|^2$ as a function
of the frequency $\omega$ is plotted. Here $M=10^2 M_{\rm{P}}$, $Q=M-e$ and
$l=413$ which is the smallest angular momentum quantum number whereby the
hole is transformed into a naked singularity in this case. 
\label{f2}}
\end{figure}

\begin{figure}[t]
\epsfig{file=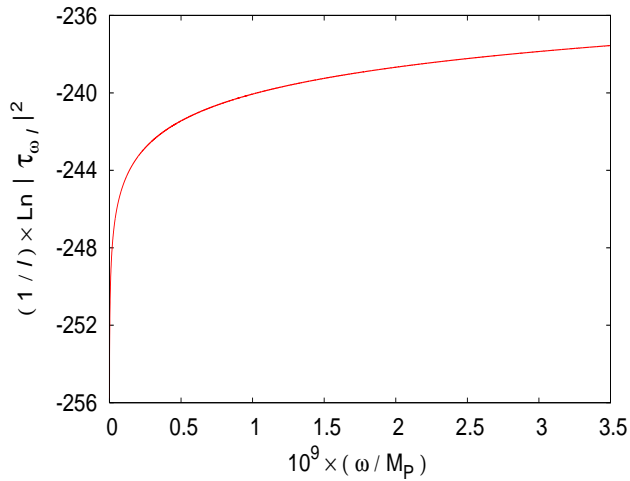,angle=0, width=8.5cm, height=6.5cm}
\caption{The absorption probability  $|{\cal{T}}_{\omega l}|^2$ as a function
of the frequency $\omega$ is plotted. Here we have chosen $M=10^5
M_{\odot}$ (no discharge condition), $Q=M-e$ and $l=1\,141\,721 \times 10^{58}$.
\label{f1}}
\end{figure}

Let us define now the ``{\it{cosmic-censorship function}}'':
\begin{equation}
{\cal{C}}(M,Q,J) \equiv M^2 (M^2-Q^2)-J^2 \label{c1}
\end{equation}
[see Eq.~(\ref{ccc})]. The initial configuration of our black hole
is given by ${\cal{C}}(M,Q,0) = M^2 (M^2-Q^2) > 0$. Now, we are able
to determine under what conditions the weak cosmic censorship
conjecture could be violated by the quantum tunneling process.
Assuming that a single particle $a^{\dagger}_{\leftarrow \omega l
m}|0\rangle$ coming from far away enters into the hole, the value of
the cosmic censorship function becomes
\begin{equation}
{\cal{C}}(M+\omega,Q,J) = {\cal{C}}(M,Q,0) - J^2 + {\cal{O}}(\omega), \label{c2}
\end{equation}
where we have required
$
\omega/M \ll {\cal{C}}(M,Q,0)/M^{4} \ll 1
$.
Thus, for particles with small enough frequencies the cosmic
censorship function will take negative values provided that
\begin{equation}
l(l+1) > {\cal{C}}(M,Q,0). \label{breaks}
\end{equation}
Although black holes near extremity are the ones which require
particles with least angular momenta to overspin, $l$ may be very
large depending on the black hole mass. For $Q=M-e$, Eq.~(\ref{breaks}) 
becomes $l(l+1) > M^3(2e-e^2/M)$. (See, e.g., Refs.~\cite{rapid,cm}
for other instances where our semiclassical approach was used with
success in the context of low frequencies and large angular 
momenta.) In Figs.~\ref{f2} and \ref{f1} we plot $|{\cal{T}}_{\omega l}|^2$ 
as a function of the particle frequency $\omega$. Our semiclassical
analysis suggests that WCCC can be quantum-mechanically violated.
Note that the absorption rate for the black hole mass $M=10^2
M_{\rm{P}}$ is much larger than the corresponding one for $M=10^5
M_{\odot}$~\cite{note}. This can be seen as an indication that perhaps any
processes which convert black holes into naked singularities should
have a quantum-mechanical nature.

In conclusion, we show how quantum effects can lead to a violation
of the weak cosmic censorship conjecture for nearly extreme
charged black holes. It seems to us
that a full quantum gravity theory should eventually play a major
role in the context of extreme black holes (see also 
Preskill {\em et al}~\cite{preskill}) either to rescue the WCCC
or to unveil the physical nature of the ``naked singularities''
which we would be forced to face. Nevertheless, we believe that as far 
as we do not have a complete theory of quantum gravity yet (where 
backreaction and other possible effects will be fully taken into 
account), the semiclassical formalism can bring useful 
new insights and may guide future research on this issue.

\begin{acknowledgments}

G.M. would like to thank J. Casti\~neiras, L. Crispino, A. Higuchi,
and D. Vanzella and  for profitable discussions. G.M.
acknowledges partial support from Conselho Nacional de
Desenvolvimento Cien\-t\'\i fico e Tecnol\'ogico and Funda\c c\~ao
de Amparo \`a Pesquisa do Estado de S\~ao Paulo and A.S.
acknowledges full support from Funda\c c\~ao de Amparo \`a Pesquisa
do Estado de S\~ao Paulo.

\end{acknowledgments}


\begin{thebibliography}{99}
\bibitem{note}
Modes with large $l$ can be thought as particles approaching the hole 
in a highly non-radial way (large impact parameter). Large $l$ values 
induce a large centrifugal barrier which cannot be overpassed classically 
by particles with low energies but still can be tunneled quantum mechanically
by them. In order to consider lower angular momenta (and, thus, centrifugal
barriers), we should take smaller values of $M$.
\end{thebibliography}
\end{document}